\documentclass{rspublic}

\usepackage{amssymb}
\usepackage{graphicx}
\begin{document}

\title[Sliding friction of elastomers]
{
State, rate and temperature-dependent sliding friction of elastomers
}

\author[O. Ronsin \& K. Labastie Coeyrehourcq]{Olivier Ronsin\footnote{To whom correspondence should be addressed} and Karine Labastie Coeyrehourcq}

\affiliation{Groupe de Physique des Solides, 2 place Jussieu 75251 Paris Cedex 05, France}

\label{firstpage}

\maketitle

\begin{abstract}{Place keywords here}
We present an experimental investigation of the non stationary frictional properties of multicontact interfaces between rough elastomers and rough hard glass at low velocities ($\leq 200\3\mu$m~s$^{-1}$). These systems, for which the deformation contribution to friction is negligible, are shown to exhibit a phenomenology which is similar to what is observed for non elastomeric materials in the same multi-contact configuration, and which are quantitatively described by the state- and rate-dependent friction laws. This permits to identify clearly the two contributions to adhesive friction which are mixed in steady sliding~: the interfacial shear stress which appears as thermally activated formation and breaking of molecular bonds, and the real area of contact which evolves through viscoelastic creep of the load bearing asperities.

\end{abstract}

\section{Introduction}
\label{sec:intro}
Studies of friction of elastomeric materials over hard surfaces have been, up to now, limited to stationary sliding, although practical situations are often non stationary.  Extensive experimental studies of the sliding friction of rubber show that the steady state dynamic friction force $F^{\textnormal{ss}}$ varies with temperature $T$ and velocity $V$ according to a single `master curve' (Grosch, 1963):
\begin{equation}
\label{eq:Fss}
F^{\textnormal{ss}} (V,T) = F^{\textnormal{ss}} (a_T V)
\end{equation}
where $a_T$ is a decreasing function of temperature, characteristic of the bulk viscoelastic properties of a given elastomer (see eg. Ferry, 1970). Furthermore (Grosch, 1963) for materials with widely different viscoelastic properties, this curve shows a single peak for smooth tracks (plane glass) with its maximum occurring at a velocity $V_{s}$ such that $V_{s}/f_{G''}$ (where $f_{G''}$ is the frequency at which the bulk loss modulus is maximum) defines a length $d_{s}$ of molecular size. For rough tracks, a second peak appears, with its maximum at a velocity $V_{h}$ such that $V_{h}/f_{\delta}$ (where $f_{\delta}$ is the frequency at which the bulk loss tangent is maximum), defines a length $d_{h}$ of the order of the track roughness. Two different mechanisms were associated with these two length scales, one attributed to adhesion loss involving interfacial molecular bonds, the other one to viscoelastic dissipation within the bulk of the asperities. For the adhesion component, the mechanism by which a frequency characteristic of the bulk might be involved is not clear, and several theories where proposed to explain the bell shaped master curve observed.

Following the seminal adhesion theory of metal friction of Bowden and Tabor (1950), Ludema \& Tabor (1966) described the adhesive friction force between a spherical hard indenter and a flat viscoelastic surface as a combination of two physical terms~: 
\begin{equation}
\label{eq:Tabor}
F = \sigma_s\Sigma_r
\end{equation}
where $\Sigma_r$ is the real area of contact and $\sigma_s$ is a mean interfacial shear stress required to break junctions. In their single Hertz contact geometry, the area of contact varies as $\Sigma_r \propto E(t_0)^{-2/3}$ where $E$ is the time dependent Young modulus evaluated at the time $t_0$ necessary for the contact to travel its own size at velocity $V$, namely $t_0 = a/V$ where $a$ is the contact size. As the viscoelastic modulus obeys a time-temperature equivalence with a shift factor $a_T$ and increases with  reduced time $t/a_T$, $\Sigma_r$ increases with $t_0/a_T$ and decreases, in stationary sliding, with $V$. The shear stress $\sigma_s$ is expected to increase with the reduced shear rate $a_T\dot\epsilon$ as observed in bulk tensile fracture experiments on the same kind of elastomers (Smith 1958).
Assuming that the shear is accommodated in a thin interfacial layer of thickness $h$, $\dot\epsilon = V/h$, and $\sigma_s$ increases with $a_T V$. Finally, the friction force reads
\begin{equation}
\label{eq:FssTaborWLF}
F^{\textnormal{ss}} (a_T V) = \sigma (a_T V)\Sigma (t_0/a_T = a/(a_T V))
\end{equation}
which produces a bell shaped curve with a position of the maximum in good agreement with experimental results, provided that $h$ is chosen to be of order $10$~nm (Ludema \& Tabor, 1966).

Savkoor (1987) used a similar approach to describe friction of rough surfaces, as the result of the statistical averaging, over the contact population, of a single micrometric contact response. He  analyzed the two effects, with a more detailed description of the evolution of the contact area, and introduced a fracture mechanics description of the shear strength, based on a Griffith-like energetic criterion for failure of the contact. This theory, though leading to the correct qualitative behavior, depends highly on details such as geometric characteristics of the surfaces.

Another approach, proposed by Schallamach (1953, 1963) focuses on the interfacial term, which is described as thermally activated formation and breaking of bonds between rubber molecules and the hard substrate. This yields a mean force necessary to break bonds that increases with velocity, but a number of bonds that decreases with velocity, leading to a global friction force that qualitatively has the observed bell shape.

In all these approaches, adhesive friction is viewed as a competition between two effects~: the material strength (force to break a bond, shear strength) increasing with velocity, and the amount of adhesive (number of bonds, area of contact) that decreases with velocity. In steady state, these effects are mixed and thus difficult to separate.

Recently, studies of non stationary friction of non elastomeric materials have been successful in separating these two contributions to the friction force. They concern friction at low velocities ($\leq 100 \mu$m s$^{-1}$) between macroscopically flat surfaces with a micrometric roughness, forming a multicontact interface (MCI) with a large but dilute population of contacts. The phenomenology of these systems, which has proved to be independent to a large extent of the nature of the materials involved (granite, paper, polymer glasses), is quantitatively described by the state- and rate- dependent friction (SRF) laws first introduced by Rice \& Ruina (1983).

The detailed analysis of polymer glasses has shown that these laws can be interpreted through the Tabor-like decomposition (\ref{eq:Tabor}) involving the real area of contact $\Sigma_r$ and the interfacial shear strength $\sigma_s$. For multicontact interfaces, this decomposition is particularly well suited to explain the Amontons-Coulomb proportionality observed between the friction force and the normal load through a linear dependence of the real area of contact on normal load, coming from the statistical distribution of asperity heights (Greenwood \& Williamson, 1966). The small variations of the friction coefficient which are responsible for the dynamical stability of the system depicted on figure~\ref{fig:setup} were extensively studied. They could be attributed to the evolution of the real area of contact with a ``state'' variable --- a time-dependent age $\phi$ --- and to the dependence of the interfacial shear strength $\sigma_s$ on the instantaneous shear rate\footnote{provided that no mechanisms of strength evolution with time, such as polymer chains interdiffusion, is at work (Berthoud {\it et al.}, 1999)}, which is itself proportional to the instantaneous velocity $\dot x$ of the slider. That is, the friction force can be written~:
\begin{equation}
\label{eq:srfa}
F(\dot x,\phi) = W\mu (\dot x,\phi) = \sigma_s (\dot x)\Sigma_r (\phi)
\end{equation}

The variations of the contact area with age could be obtained by measuring the increase in static friction with the time of contact $t_c$ prior to sliding. For polymer glasses, this is associated to the plastic creep of the load-bearing asperities, as observed by direct visualization of the real area of contact (Dieterich \& Kilgore, 1994) and by quantitative comparison with the bulk plastic creep properties for various temperatures below the glass transition (Berthoud {\it at al.}, 1999). In steady sliding, this ageing occurs during a time $D_0/V$ where $D_0$ is a characteristic length found to be of the same order of magnitude as the mean contact size. The friction force then has a form similar to Ludema \& Tabor's~:
\begin{equation}
\label{eq:FssTabor}
F^{\textnormal{ss}} (V) = \sigma_s (V)\Sigma_r (D_0/V)
\end{equation}
which, at low velocity ($\leq 100\ \mu$m$\3$s$^{-1}$), was found to be a decreasing function of velocity, leading to the existence of a bifurcation between steady sliding and stick-slip oscillations. This bifurcation could be analyzed quantitatively with a time evolution law for the state variable $\phi$ in non steady conditions that accounts for the past history of the system over the length $D_0$, and that interpolates between the static and steady state regimes~:
\begin{equation}
\label{eq:ageint}
\phi (t) = \int_{-\infty}^{t} \exp \left(-\frac{x(t)-x(t^{\prime})}{D_0}\right)\ dt^{\prime}
\end{equation}
which takes the differential form~:
\begin{equation}
\label{eq:srfb}
\dot \phi = 1-\frac{\dot x \phi}{D_0}
\end{equation}
The dependences of real area of contact on age and of shear strength on velocity were found to be logarithmic~:
\begin{eqnarray}
\label{eq:rheoglass}
\sigma_s (\dot x) & = & \sigma_s^0 \left(1+\alpha\ln (\dot x/V_0)\right)\\
\Sigma_r (\phi) & = & \Sigma_r^0 \left(1+m\ln (\phi/\phi_0)\right)
\end{eqnarray}
with values of $\alpha$ and $m$ small compared to unity. The friction coefficient thus reduces, to first order in $\alpha$ and $m$, to the usual SRF expression~:
\begin{equation}
\label{eq:srflog}
\mu (\dot x,\phi) = \mu_0 + A\ln\left(\frac{\dot x}{V_0}\right) + B\ln\left(\frac{\phi}{\phi_0}\right)
\end{equation}
When the steady state friction decreases with velocity (i.e. when $B-A >0$), this friction laws predicts (see \ref{sec:appendix}) a bifurcation from steady sliding to stick-slip oscillations at low system stiffness and/or high normal loads for the system depicted on figure~\ref{fig:setup}, as observed experimentally. The detailed study of this bifurcation, coupled with steady state and static measurements, gives access to all the parameters involved in the SRF laws. This kind of analysis, performed as a function of temperature for polymer glasses, showed that the shear response of the contacts, which is expected to occur at the junction between asperities within a molecular thick layer of weaker material (confined polymer chain ends) is an activated process with an activation volume of order (1 nm)$^3$, indicating that this nanometer thick layer between the contacts slides by activated depinning of uncorrelated nanometric blocks, giving a picture very similar to Schallamach's description of interfacial sliding of rubbers (Schallamach, 1953, 1963).

This paper deals with the extension of this description of non stationary friction to elastomers. We show that the low velocity friction properties of a multicontact interface between elastomers and rough hard glass, studied as a function of  temperature, are indeed quantitatively described by the SRF laws. The viscoelastic properties of the material appear through a temperature dependence of the SRF parameter $A$ and $B$ compatible with a WLF transform.
This permits to separate clearly the shear strength and the area of contact contributions to the friction force, and to bridge between Tabor and Schallamach's descriptions of elastomer friction. The paper is organized as follows~: in the next section, we describe the experimental setup. Section~\ref{sec:results} presents the results of the static, steady state sliding and unsteady friction measurements, followed by a discussion of these results within the SRF framework.

\section{Experimental}
\label{sec:experimantal}
\subsection{Samples}

We studied three elastomers made of synthetic poly-isoprene, reticulated with dicumyl peroxide at $443$ K. The glass transition temperature $T_{\textnormal{g}}$, determined by differential scanning calorimetry, was changed by adding plasticisers (liquid paraffin or resin). The samples were moulded in $2.8$ mm thick plates. The composition in mass and the glass transition temperatures of the samples are shown on table~\ref{tab:samples}.
\begin{table}
\caption{\label{tab:samples}Sample composition and glass transition temperature}
\begin{center}
\begin{tabular}{lccc}
\hline
 & \multicolumn{1}{c}{Sample \#1} & \multicolumn{1}{c}{Sample \#2}& \multicolumn{1}{c}{Sample \#3} \\
\hline
poly-isoprene & $100$ & $100$ & $100$ \\
paraffin & $20$ & --- & --- \\
resin & --- & $27$ & --- \\
peroxide & $1.2$ & $1.2$ & $1.2$ \\
\hline
glass transition (K) & $303$ & $296$ & $274$ \\
\hline
\end{tabular}
\end{center}
\end{table}

The dynamic mechanical properties (storage $G^{\prime}$ and loss moduli $G^{\prime\prime}$) of the three elastomer samples were obtained in the frequency range $0.1 - 750$ Hz for various temperatures. This permitted to perform a WLF transform (shown on figure~\ref{fig:at}(a) for sample \#1), leading to the shift factors $a_T$ that are presented, on figure~\ref{fig:at} with the glass transition as a reference temperature. Most of the results obtained are, unless stated, qualitatively independent of the sample and we will therefore present the quantitative results obtained on sample \#1.

The slider is composed of an aluminum plate ($10\times10\times1$ cm$^3$) on which was glued the elastomer sample ($7\times 7\times 0.28$ cm$^3$). The surface of the slider was hand lapped at low temperature (below $T_{\textnormal{g}}$) with a SiC abrasive powder of nominal grain size $23\3\mu$m. This procedure has proved to give a reproducible rms roughness of $1.3\3\mu$m.

The track ($15\times 15\times 1$ cm$^3$) is made of rough glass prepared in the same way. We use a glass track in order to avoid any dependence of the shear strength $\sigma_s$ with the contact age $\phi$ that might be caused by inter-diffusion of polymer chains.

\subsection{Friction apparatus}

The details of the experimental setup sketched on figure~\ref{fig:setup} are described in detail elsewhere (Berthoud \textit{et al.} 1999). The slider is driven along the track by a translation stage, composed of a stepping motor pushing a cantilever spring of stiffness $k$. The spring pushes the slider through a cross-cylinder junction in such way that the contact is in the same plane as the track-slider interface. The velocity range of the driving stage is $V = 0.1$ to $200$ $\mu$m$\3$s$^{-1}$ ensuring that frictional heating is negligible. The spring stiffness $k$ was chosen to be the most compliant part of the setup in order to be a control parameter of the stability of the system. Using the relaxed storage modulus of the most compliant sample, we can estimate a lower limit for the shear stiffness of the samples in the temperature range studied of $10^6$ N$\3$m$^{-1}$. We used cantilever springs of maximal stiffness $2.4\times10^5$ N$\3$m$^{-1}$. The force applied to the slider is deduced from the spring deflection measured by an eddy current inductive transducer with a sub-micron resolution within the frequency range.

The temperature control of the samples was achieved by gluing resistive stripes (total of $140$ W) under the track and over the slider, each with a $100 \Omega$ platinum probe and a controller for thermal regulation. The elastomer temperature was also measured with a separate probe located inside it. Both track and sample where enclosed in a styrofoam box. The temperature was thus controlled  between room temperature and $370$ K with $0.5$ K precision. It was found that humidity has an influence on the results only when above $60\%$ and the results presented here were obtained under dryer conditions.

\subsection{Friction force measurement}

This setup allows the measurement of both static and dynamic friction forces. The static force $F_s$ is defined from the peak value of the tangential force when loading rapidly the slider from rest.
Both static and dynamic friction forces show similar fluctuations that depend on the slider's position along the track and which correspond to long wavelength inhomogeneities of the interface. They can thus be removed by measuring the friction forces $\Delta F$ relative to dynamic friction in steady state at a reference velocity $V_{\textnormal{ref}}$ measured at the same position. This enabled us to measure the variations of friction forces with an error lower than the track fluctuations.
The absolute value of the force is obtained by adding to $\Delta F$ the mean value of the friction force at the reference velocity $V_{\textnormal{ref}}$ over the whole track.

Figure~\ref{fig:AC} shows the static and dynamic friction forces measured as a function of the normal load $W$.

The proportionality defines the static and dynamic friction coefficients as the slopes $\mu_s = F_s/W$ and $\mu_d = F_d/W$. This proportionality of the friction force to the normal load is usually attributed to a linear relation between the real area of contact and the normal load, as expected from Greenwood's description of multicontact interfaces (Greenwood \& Williamson, 1966).

\section{Results}
\label{sec:results}

\subsection{Static friction}

In order to study the dependence of the static friction force on contact time $t_c$ prior to sliding, we follow a procedure proposed by Berthoud \textit{et al.} (1999) : the slider is first put in steady sliding at a velocity $V_{\textnormal{ref}}$ during $10$ seconds in order to renew the contact population. It is then stopped, remaining under tangential load for a resting time $t_c$. After that time, the slider is shear-loaded at velocity $V_{\textnormal{load}}$ and the force is recorded. The static friction is then defined from the peak value of the force, at which point the instantaneous velocity of the slider equals the loading velocity. It has been shown (Berthoud \textit{et al.} 1999) that the loading rate has an effect on the value of the static friction coefficient. We thus measure it at the constant loading rate $V_{\textnormal{load}} = 100 \mu$m$\3$s$^{-1}$. For convenience, the reference velocity $V_{\textnormal{ref}}$ is chosen equal to the loading velocity.
Figure \ref{fig:mus} shows the static friction coefficient $\mu_s$ as a function of contact time $t_c$ for different temperatures.
We find a quasi-logarithmic increase of the static friction with rest time, with a mean slope depending strongly on temperature. In order to analyze this dependence within the SRF framework, we define a local logarithmic slope $\beta_s = d\mu_s/d\ln t_c$ over the first two decades of time.
Figure \ref{fig:betas} shows the dependence of $\beta_s$ on $T$, showing a clear maximum around $60^{\circ}$ C. 
In order to analyse the influence of temperature on this ageing effect, we tried to check whether it obeys a time-temperature equivalence. By applying a temperature-dependent scaling $1/a_T^s$ in time to each set of measurements, we could obtain a single `master curve' (figure~\ref{fig:muswlf}(a)) for the static friction coefficient~:
\begin{equation}
\label{eq:muswlf}
\mu_s (t_c, T) = \mu_s (t_c/a_T^s)
\end{equation}
Furthermore, figure~\ref{fig:muswlf}(b) shows that the shift factors $a_T^s$ agree quantitatively with the viscoelastic bulk ones. Static ageing is thus the result of an increase of the real area of contact through viscoelastic creep of the load bearing asperities.

\subsection{Steady state dynamic friction}

The steady state dynamic friction coefficient $\mu_d^{\textnormal{ss}} (V)$ was measured for various temperatures, as shown on figure~\ref{fig:mud}.
The friction always decreases with velocity within the temperature range studied, but with slopes that depend on temperature. Again, we define a local logarithmic slope $\beta_d = -d\mu_d^{\textnormal{ss}}/d\ln V$ between $0.1$ and $10$ $\mu$m$\3$s$^{-1}$, which is shown as a function of temperature on figure~\ref{fig:betad}. It exhibits a clear maximum at about $330$ K and the trend at higher temperatures indicates that $\beta_d$ might become negative.
As for static friction, we are able to define temperature-dependent scaling factors $a_T^d$ in time, with the help of which measurements at various temperatures fall on a single `master curve' shown on figure~\ref{fig:mudwlf}(a)~:
\begin{equation}
\label{eq:mudwlf}
\mu_d^{\textnormal{ss}} (V,T) = \mu_d^{\textnormal{ss}} (a_T^d V)
\end{equation}
Figure~\ref{fig:mudwlf}(b) shows that the scaling factors $a_T^d$ compare quantitatively with bulk ones, in agreement with previous sliding friction measurements on various elastomers (Grosch, Ludema \& Tabor), although we don't see the peak that was observed in these experiments. However, the saturation found at low reduced velocity indicates that such a peak might be present in our system at higher temperatures.

\subsection{Non stationary friction, stick-slip bifurcation}

Because of the velocity weakening behavior of the steady friction force, stationary sliding could only be obtained under low loads and with stiff external springs. The system otherwise slides through a succession of sticking and slipping phases. The bifurcation between these two dynamical regimes is continuous : the amplitude of the stick-slip oscillations continuously grows from zero when increasing the normal load as shown on figure~\ref{fig:bif} (a).

Figure \ref{fig:bif} (b) shows the dynamical phase diagram of the system in the ($k/W$, $V$) space, at room temperature. The critical reduced stiffness slightly depends on the velocity $V$ as observed on other materials in the same multicontact configuration.

This kind of bifurcation is accounted for by the linear stability analysis of the SRF laws (see~\ref{sec:appendix}). It is predicted to occur for
\begin{equation}
\label{eq:bif}
\left(\frac{k}{W}\right)_c = \frac{B-A}{D_0} = \frac{\beta_d}{D_0}
\end{equation}

The critical reduced stiffness $(k/W)_c$ was measured at a driving velocity $V = 2 \mu$m\3s$^{-1}$ for various temperatures, together with the pulsation $\Omega_c$ of the stick-slip oscillations at the bifurcation. Using the previous measurements of the velocity weakening slope $\beta_d$ at corresponding temperatures, we can then deduce the value of the memory length $D_0 = \beta_d/(k/W)_c$.

Figure \ref{fig:D0} shows the values obtained for different temperatures for the three elastomers used. Though the dispersion is large due to cumulative errors on $\beta_d$ and $(k/W)_c$, the memory length appears independent of temperature and sample type, having an average value of $D_0\simeq 1\3\mu$m, comparable with the surface roughness, in accordance with the geometrical interpretation of this length in terms of a slip distance necessary to renew the contact population.

The pulsation $\Omega_c$ gives access to the dependence of the friction coefficient on the instantaneous velocity $A = \partial\mu_d /\partial {\ln\dot x}$ as (see \ref{sec:appendix})
\begin{equation}
\label{eq:A}
A = \left(\frac{k}{W}\right)_c\frac{V^2}{D_0 \Omega_c^2}
\end{equation}

Its temperature dependence is shown on figure~\ref{fig:A}.
The indirectness of the measurement leads to cumulative errors that become large, but there is nevertheless a clear increase with temperature.

\section{Discussion}
\label{sec:discussion}
We have so far shown that the velocity weakening regime in steady sliding leading to a continuous bifurcation between steady sliding and stick-slip oscillations is compatible with the phenomenological state- and rate-dependent friction laws. This permitted the determination of the various parameters involved. A further step can be made by trying to link the dynamic response to the static threshold. Provided that the loading phase is short compared to the static age $t_c$ of the multicontact interface, static friction is a transient measurement at the loading velocity $V_{\textnormal{load}}$ with an age $t_{c}$, and following the SRF description, the static friction coefficient can be written as
\begin{equation}
\label{eq:musvload}
\mu_s (t_c) = \mu_0 + A\ln (V_{\textnormal{load}}/V_0)+B\ln(t_c/t_0)
\end{equation}
This implies a logarithmic dependence of the static friction on the loading velocity, which was indeed observed and found quantitatively compatible with this law for polymer glasses (Berthoud {\textit{et al.}} 1999). Furthermore, the logarithmic ageing slope $\beta_s$ should compare with $B$. As we have measured independently the direct effect parameter $A$, and the steady state friction slope $\beta_d = B-A$, the comparison can be done as shown on figure~\ref{fig:statdyn} where $\beta_d+A$ and $\beta_s$ are plotted for different temperatures. 
The agreement indicates that the state-and rate-dependent friction model coherently describes the friction of rough elastomers on rough glass. Furthermore, the memory length $D_0$ was found to be independent of either temperature or material and to be on the order of the mean contact size. This shows that the interpretation of the age parameter in terms of real area of contact which was shown to hold for polymer glasses also holds here, though the physical origin of the ageing is clearly viscoelastic in our case, as was suggested by Ludema \& Tabor (1966). This is confirmed by the fact that the dependence of static friction on both time and temperature follows a time-temperature equivalence compatible with bulk viscoelastic measurements.

The comparison of our results with other experimental studies of steady sliding friction on elastomers is limited because, with our system, we only observed velocity weakening behavior. However, we have seen that the saturation of $\mu_d^{\textnormal{ss}}$ at low reduced velocities (figure~\ref{fig:mudwlf}(a)) may be the trace of the maximum observed in previous investigations. We assumed that the viscoelastic dissipation due to the deformation of asperities while sliding was negligible compared to adhesive dissipation. This can be checked by an evaluation of the deformation contribution to the friction coefficient as a function of viscoelastic properties of the material and the geometric characteristics of the surfaces forming the multicontact interface. At sliding velocity $V$, the contacts are stressed during a time of the order of $V/a$ where $a$ is the mean contact size. Such a contact will be subjected to a deformation $\epsilon \simeq a/R$, where $R$ is the asperity radius of curvature, within a volume of order $a^3$. The energy dissipated by viscoelastic losses in the contact will then be of order
\begin{equation}
\label{eq:dissipnrj}
E_{\textnormal{def}} \simeq G^{\prime\prime} (V/a) \times \epsilon^2 \times a^{3} \simeq G^{\prime\prime} (V/a) \frac{a^5}{R^2}
\end{equation}
This dissipation will result in a contribution $F_{\textnormal{def}}$ to the friction force such that $a F_{\textnormal{def}} = E_{\textnormal{def}}$, or to a contact shear resistance
\begin{equation}
\label{eq:sheardef}
\sigma_{\textnormal{def}} \simeq G^{\prime\prime}\frac{a^2}{R^2}
\end{equation}
From Greenwood's (Greenwood \& Williamson, 1966) description of elastic multicontact interfaces, we can estimate the mean contact size as ${\bar a}\simeq \sqrt{R\sigma}$ and the mean contact pressure as
\begin{equation}
\bar p = W/\Sigma_r = G^{\prime}\left(\frac{\sigma}{R}\right)^{1/2}
\end{equation}
where $\sigma$ is the characteristic length of the asperity height distribution. The deformation loss will thus give a contribution to the friction coefficient of the order of
\begin{equation}
\label{eq:mudef}
\mu_{\textnormal{def}}(a_T V) \simeq \frac{\sigma_{\textnormal{def}}}{\bar p} \simeq \left(\frac{\sigma}{R}\right)^{1/2}\left(\tan \delta\right)_{a_T f = V/\sqrt{\sigma R}}
\end{equation}
This is in agreement with the experiments of Grosch (1966) on dry silicon carbide paper for which $\sigma$ and $R$ are expected to be of the same order of magnitude. Note that a similar result was obtained by Persson (1998) but in the limited case of very rough surfaces for which $\sigma$ and $R$ are of the same order of magnitude. For our surfaces, these quantities, measured from surface profiles, are of order $\sigma \simeq 1\3\mu$m and $R\simeq 100\3\mu$m. Furthermore, the loss tangent of sample \#1 has a maximum of $0.9$ about a reduced frequency $a_{T}f_{\delta}\simeq 10^{-3}$ Hz. The deformation component of friction is thus expected to have a maximum of order $0.09$ at a reduced velocity $a_T V\simeq a_T f_\delta\sqrt{\sigma R} \simeq 10^{-2}\3\mu$m$\3$s$^{-1}$. From figure~\ref{fig:mudwlf}, we see that there is no peak about this velocity and that the measured friction is about five times higher. This is also the case for the other samples and the observed friction comes clearly from adhesive mechanisms in the contact junctions. From our interpretation of this adhesive component of friction, the change from a regime of velocity weakening to velocity strengthening at high temperatures or low velocities would arise from an inefficient ageing (state) effect compared to the rheological shear response (rate) of the interface. This gives further support to the adhesion view of Ludema \& Tabor. We thus propose an extension of their description of elastomer friction accounting for non steady dynamics as
\begin{equation}
\label{eq:Tabor1}
F (\dot x,\phi,T) = \sigma_s (a_T\dot x)\Sigma_r (\frac{\phi}{a_T})
\end{equation}
with the time evolution of the state variable $\phi$ given by equation~(\ref{eq:srfb}), which reduces to  equation~(\ref{eq:FssTaborWLF}) in steady state, the contact size being averaged over the contact population. We don't expect the interfacial shear strength to compare with the bulk shear strength of the elastomer as no wear effect could be observed. A mechanism similar to Schallamach's (1953) description of interfacial friction, in terms of thermally activated and uncorrelated depinning of interfacial molecular bonds, as was proposed for glassy polymers, is more probable.

\section{Conclusion}
\label{sec:conclusion}

We have shown that the non steady adhesive friction of elastomers over hard glass, with a well defined surface geometry forming a multicontact interface is quantitatively described by the phenomenological state- and rate-dependent friction laws. This permits the separation between the two contributions to the friction force~: the interfacial shear strength and the area of contact, which are mixed in steady sliding. This extends Ludema \& Tabor's (1966) description of elastomers friction in terms of an interfacial shear strength similar to that invoked by Schallamach (1953, 1963), and a real area of contact that was shown to evolve through viscoelastic creep within the bulk of the contacting asperities leading to memory effects over a length characteristic of the mean contact size.

\begin{acknowledgements}
We are grateful to T. Baumberger and Ch. Caroli for fruitful discussions and critical reading of the manuscript. The Groupe de Physique des Solides is ``associ\'e au Centre de la Recherche Scientifique et aux Universit\'es Paris 6 et Paris 7''. 
\end{acknowledgements}

\appendix{}
\label{sec:appendix}
This appendix presents the linear stability analysis of steady sliding at velocity $V$ of the dynamical system shown on figure~\ref{fig:setup} with a friction force between the slider and the track described by a friction coefficient depending on the instantaneous velocity $\dot x$ of the slider, and on the state variable $\phi$, the time evolution of which is given by equation~(\ref{eq:srfb}). At the low velocities considered, inertia can be neglected and the dynamical equations of the system are :
\begin{eqnarray}
\label{eq:dyneq}
-W\mu_d ({\dot x},\phi) + k(x_0+Vt-x) = 0\\
{\dot\phi} = 1-\frac{{\dot x}\phi}{D_0}
\end{eqnarray}
where $x_0$ is the spring elongation in steady sliding at the driving velocity, for which
\begin{eqnarray}
\label{eq:statsol}
{\dot x} & = & V\\
\phi & = & \frac{D_0}{V}\\
\mu_d^{\textnormal{ss}}(V) & = & \mu_d (V,D_0/V)
\end{eqnarray}
We are looking for the dynamical evolution of small perturbations about the steady state solution
\begin{eqnarray}
\label{eq:pert}
\dot x & = & V+{\delta\dot x}\\
\phi & = & \frac{D_0}{V}+\delta\phi
\end{eqnarray}
with ${\delta\dot x}\ll V$ and $\delta\phi\ll D_0/V$
Substituting into the dynamical equations, and expanding to first order about the steady state solution, we find
\begin{eqnarray}
\label{eq:dynlin}
\frac{\partial \mu_d}{\partial\dot x}{\delta\dot x}
+\frac{\partial \mu_d}{\partial\phi}\delta\phi
+\frac{k}{W}\delta x = 0\\
{\delta\dot\phi} = -\frac{{\delta\dot x}}{V}
-\frac{V}{D_0}\delta\phi
\end{eqnarray}
where partial derivatives are evaluated at the steady state value ($\dot x = V, \phi = D_0 /V$). Looking for solutions of the form
\begin{eqnarray}
\label{eq:solform}
\delta x & = & \delta x_0 e^{st}\\
\delta\phi & = & \delta\phi_0 e^{st}
\end{eqnarray}
leads to a linear system of two equations in $\delta x_0$ and $\delta\phi_0$ that has non trivial solutions if
\begin{equation}
\label{eq:stabeq}
s^2\frac{\partial\mu_d}{\partial\dot
x}+s\left[\frac{k}{W}+\frac{V }{D_0}\frac{d\mu_d^{\textnormal{ss}}}{d V}\right]+\frac{kV}{D_0 W} = 0
\end{equation}
where we have used the relation
\begin{equation}
\label{eq:statrel}
\frac{d\mu_d^{\textnormal{ss}}}{d V} = \left(\frac{\partial\mu_d}{\partial\dot x}
-\frac{D_0}{V^2}\frac{\partial\mu_d}{\partial\phi}\right)_{\substack{\dot x = V\\ \phi = D_0/V}}
\end{equation}
Introducing
\begin{eqnarray}
\label{eq:locparam}
B-A & = & -\frac{d\mu_d^{\textnormal{ss}}}{d\ln V}\\
A & = & \left.\frac{\partial\mu_d}{\partial\ln \dot x}\right|_{\substack{\dot x = V\\ \phi = D_0/V}}
\end{eqnarray}
which might here depend on $V$, whereas in the usual SRF model are assumed to be constant, equation (\ref{eq:stabeq}) reads
\begin{equation}
\label{eq:statrelsrf}
s^2 \frac{A}{V}+s\left[\frac{k}{W}-\frac{B-A }{D_0}\right]+\frac{kV}{D_0 W} = 0
\end{equation}
The stability of the solutions (\ref{eq:solform}) are given by the sign of the real part of the roots of this equation. A positive real part leads to an amplification of the perturbation an to instability of the steady sliding solution, whereas a negative real part implies the attenuation of the perturbation and steady sliding is stable.
If $A < 0$, equation~(\ref{eq:statrelsrf}) has two real roots of opposite sign~:
\begin{equation}
\label{eq:realroot}
s_{\pm} = -\frac{V}{2A}\left[\frac{k}{W}-\frac{B-A }{D_0}\right]\pm \frac{V}{2A}\sqrt{\left[\frac{k}{W}-\frac{B-A }{D_0}\right]^2-4\frac{Ak}{D_0 W}}
\end{equation}
the positive one leading to an unstable solution (\ref{eq:solform}), and steady sliding is always unstable. If $A > 0$, the two roots are complex conjugate~:
\begin{equation}
\label{eq:complexroot}
s_{\pm} = -\frac{V}{2A}\left[\frac{k}{W}-\frac{B-A }{D_0}\right]\pm \frac{iV}{2A}\sqrt{4\frac{Ak}{D_0 W}-\left[\frac{k}{W}-\frac{B-A }{D_0}\right]^2}
\end{equation}
a bifurcation happens when the real part of $s_{\pm}$ changes sign, steady sliding is stable if
\begin{equation}
\label{eq:thres}
\frac{k}{W} > \left(\frac{k}{W}\right)_c = \frac{1}{D_0}(B-A)
\end{equation}
and unstable otherwise.
Furthermore, the unstable solution is oscillating at a pulsation given by the imaginary part of the roots $s_{\pm}$, which reads, at the bifurcation~:
\begin{equation}
\label{eq:puls}
\Omega_c = V\sqrt{\left(\frac{k}{W}\right)_c/AD_0}
\end{equation}

\newpage

\begin{figure}
\caption{\label{fig:setup}Sketch of the experimental setup : multicontact interface between nominally flat surfaces with a micrometric roughness. Slider = elastomer, track = glass. The slider, under normal load $W$ is pulled by a spring of stiffness $k$, whose end is driven at constant velocity $V$.}
\end{figure}

\begin{figure}
\caption{\label{fig:at}(a) Storage $G^{\prime}$ and loss moduli $G^{\prime\prime}$ as a function of reduced frequency $a_T f$ for sample \#1 ($T_{\textnormal{g}} = 303$ K). (b) Shift factors $a_T$ with the glass transition as a reference temperature for the three samples ($\square$ sample \#1; $\triangle$ sample \#2; $\circ$ sample \#3).}
\end{figure}

\begin{figure}
\caption{\label{fig:AC}Static friction force ($\bullet$) after a resting time of $t_c = 120$ s, and steady state dynamic friction force ($\circ$) at velocity $V = 100\3\mu$m$\3$s$^{-1}$, as a function of the normal load $W$ for sample \#1 at room temperature. The constant slope defines the static and dynamic friction coefficients $\mu_s$ and $\mu_d^{\textnormal{ss}}$.}
\end{figure}

\begin{figure}
\caption{\label{fig:mus}Static friction coefficient of sample \#1 as a function of the resting time $t_c$ for three temperatures ($\circ$ 293 K; $\times$ 333 K; $\blacksquare$ 368 K). At not too large times, the static ageing effect is clearly logarithmic, and the slope depends strongly on temperature.}
\end{figure}

\begin{figure}
\caption{\label{fig:betas}Logarithmic slope $\beta_s$ of the static ageing in the range $10$ -- $10^3$ s, as a function of temperature relative to the glass transition temperature (sample \#1, $T_{\textnormal{g}} = 303$ K).}
\end{figure}

\begin{figure}
\caption{\label{fig:muswlf}(a) Static friction coefficient of sample \#1 as a function of the reduced contact time $t_c/a_{T}^s$, with scaling factors $a_T^s$ computed to obtain a single curve ($\circ$ 293 K; $\square$ 313 K; $\diamond$ 323 K; $\times$ 333 K; $+$ 338 K; $\triangle$ 343 K; $\bullet$ 353 K; $\blacksquare$ 368 K). (b) Comparison between these shift factors ($\bullet$) and those obtained from bulk viscoelastic measurements ($\times$).}
\end{figure}

\begin{figure}
\caption{\label{fig:mud}Dynamic friction coefficient $\mu_d^{\textnormal{ss}}$ as a function of the steady velocity $V$ for three different temperatures~: $\circ$ 293 K; $\diamond$ 333 K; $+$ 369 K.}
\end{figure}

\begin{figure}
\caption{\label{fig:betad}Local logarithmic slope $\beta_d$ of the velocity dependence of the friction coefficient as a function of temperature.}
\end{figure}

\begin{figure}
\caption{\label{fig:mudwlf}(a) Steady state dynamic friction coefficient as a function of the reduced velocity $a_T^d V$ (reference temperature $T_{\textnormal{g}}$). The figure shows the measurements at different temperatures $\circ$ 293 K; $\square$ 308 K; $\diamond$ 333 K; $\times$ 353 K; $+$ 368 K; (b) Shift factors from (a) $\bullet$ compared with bulk ones $\times$ (figure~\ref{fig:at}).}
\end{figure}

\begin{figure}
\caption{\label{fig:bif}(a) Friction force as a function of time for a driving velocity of $V = 2\3\mu$m$\3$s$^{-1}$ for different values of the normal load. The steady state sliding is observed at low loads, whereas stick-slip oscillations occur at higher loads, with an amplitude that continuously increase from zero above the critical load. (b) stability diagram of the system as a function of driving velocity $V$ and stiffness over normal load $k/W$.}
\end{figure}

\begin{figure}
\caption{\label{fig:D0}Memory length $D_0$ measured from the steady-sliding --- stick-slip bifurcation at different temperatures for the three elastomer samples ($\bullet$ sample \#1; $\square$ sample \#2; $\diamond$ sample \#3).}
\end{figure}

\begin{figure}
\caption{\label{fig:A}Direct effect parameter $A$ measured from the bifurcation at $V = 2\mu$m.s$^{-1}$ as a function of temperature.}
\end{figure}

\begin{figure}
\caption{\label{fig:statdyn}Comparison between independent measurements of the ageing component of the friction~: from static friction $\beta_s$ ($\times$) and dynamic friction $\beta_d+A$ ($\circ$).}
\end{figure}
\vfill
\label{lastpage}
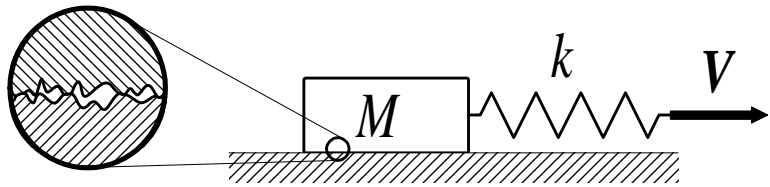
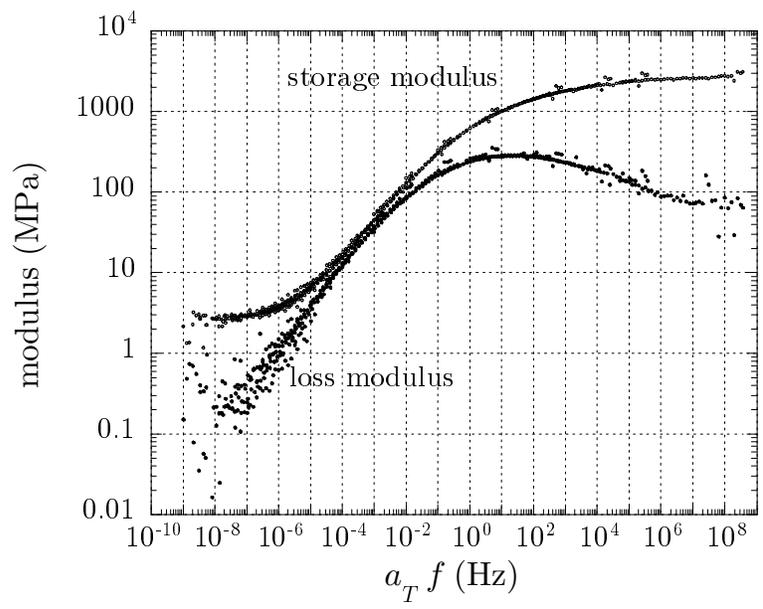
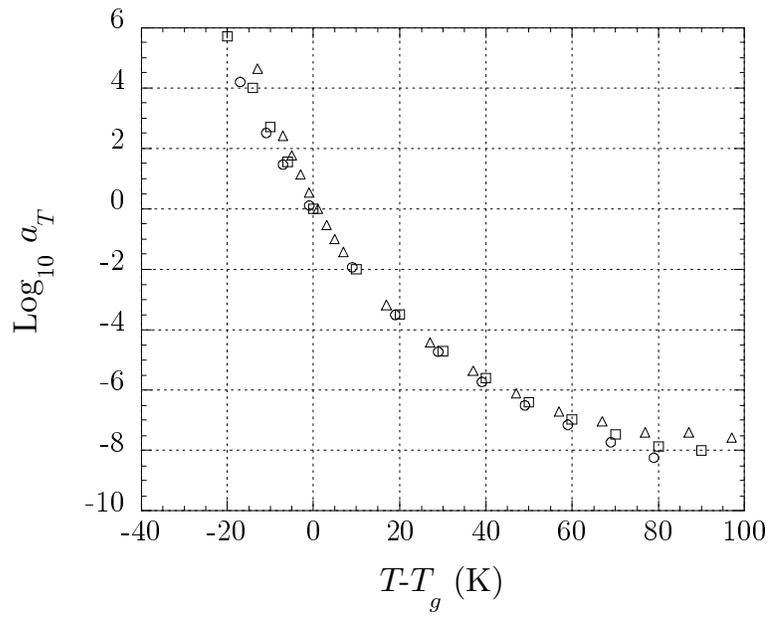
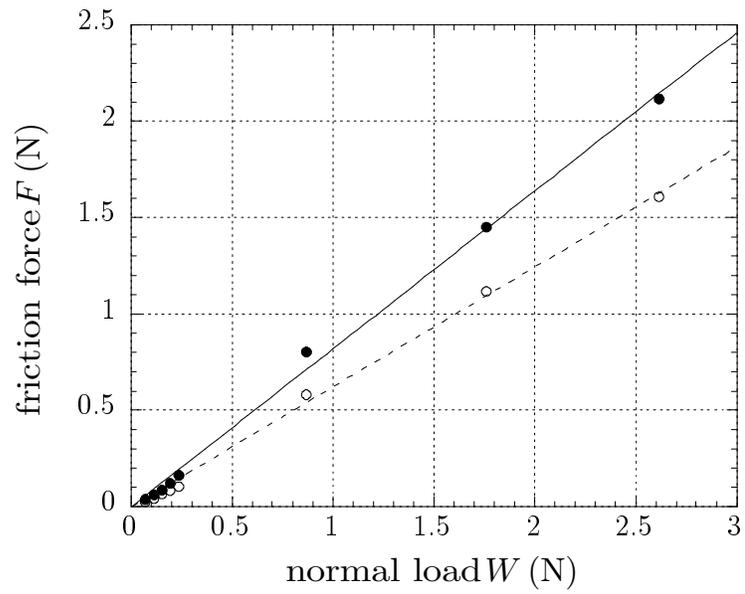
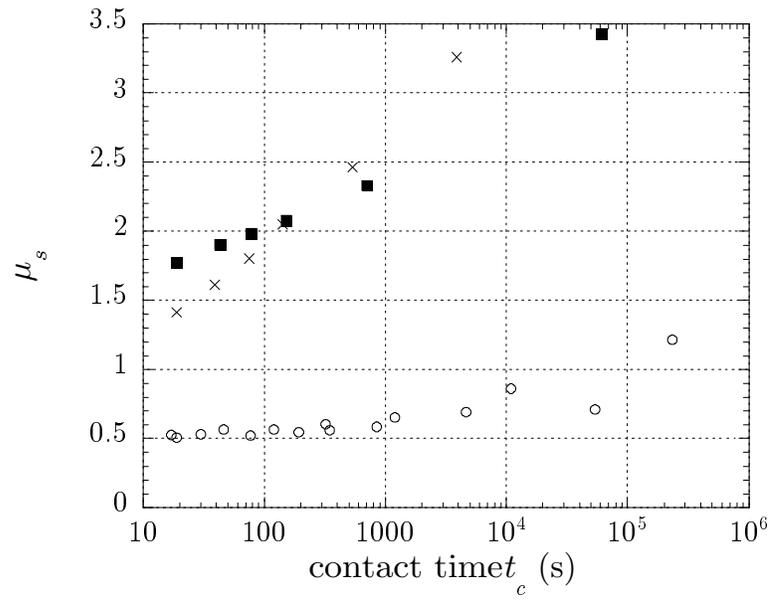
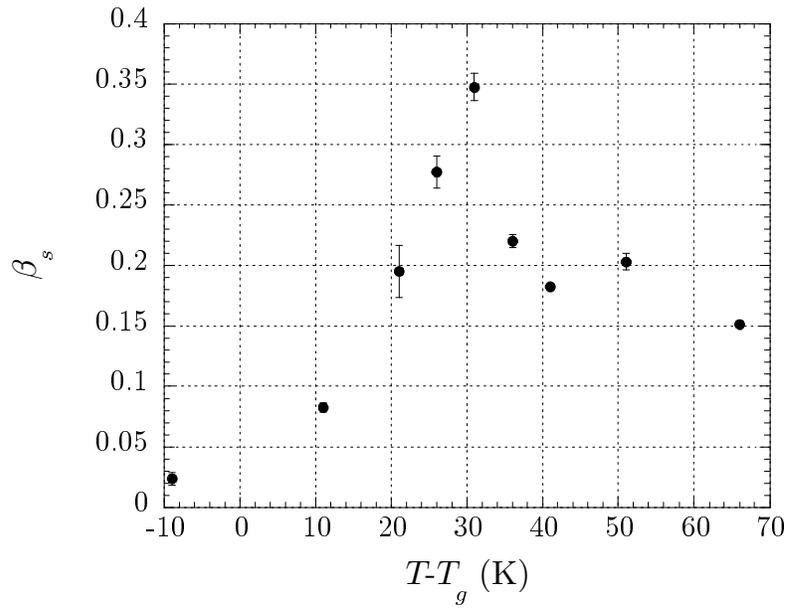
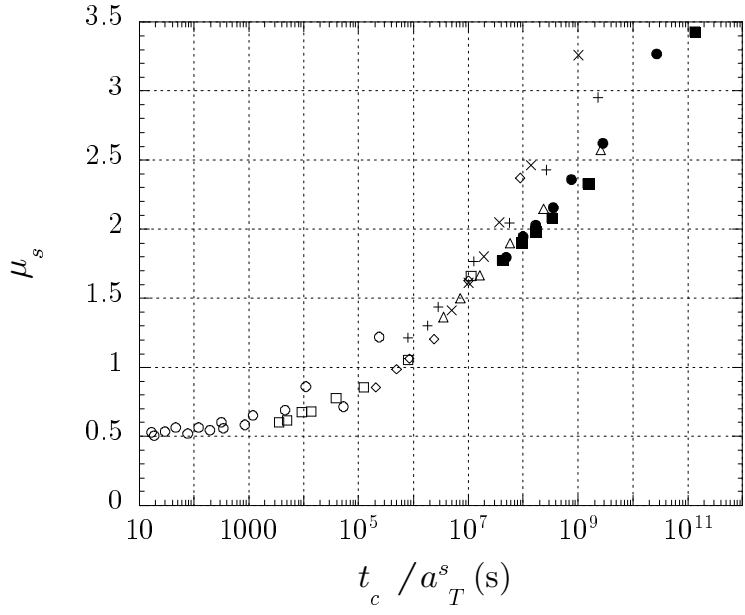
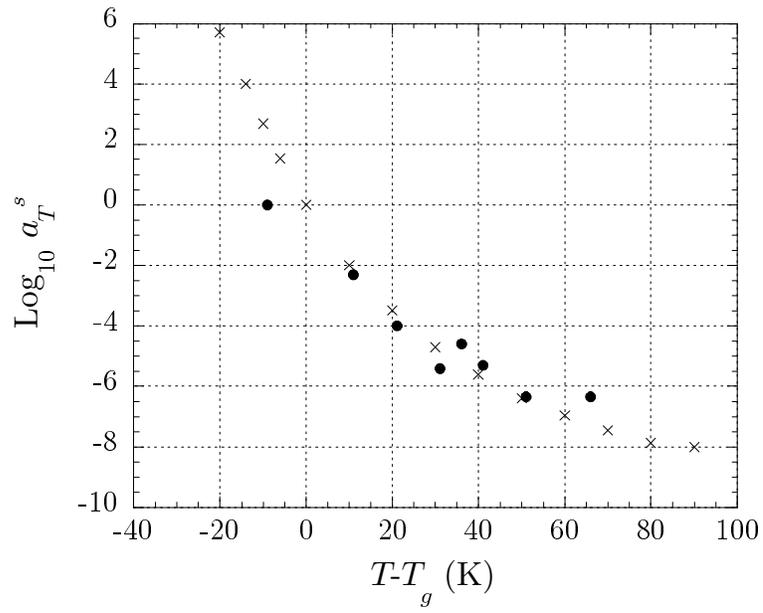
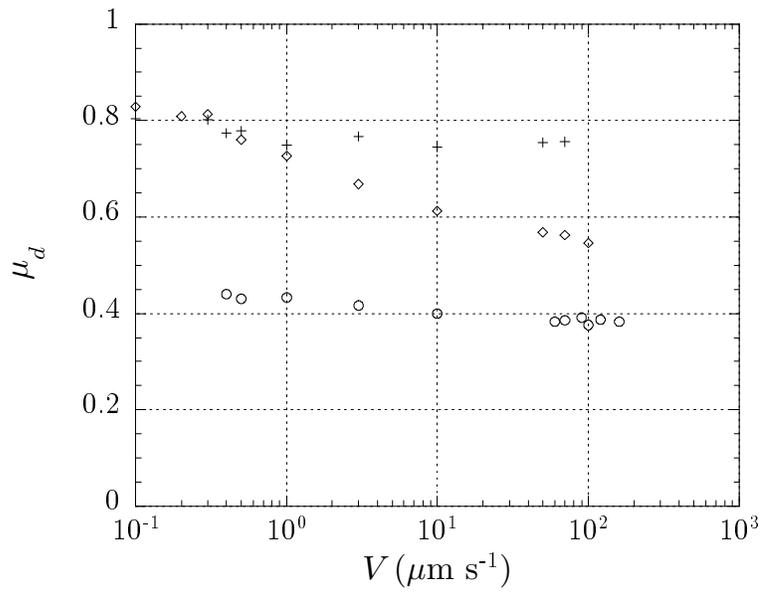
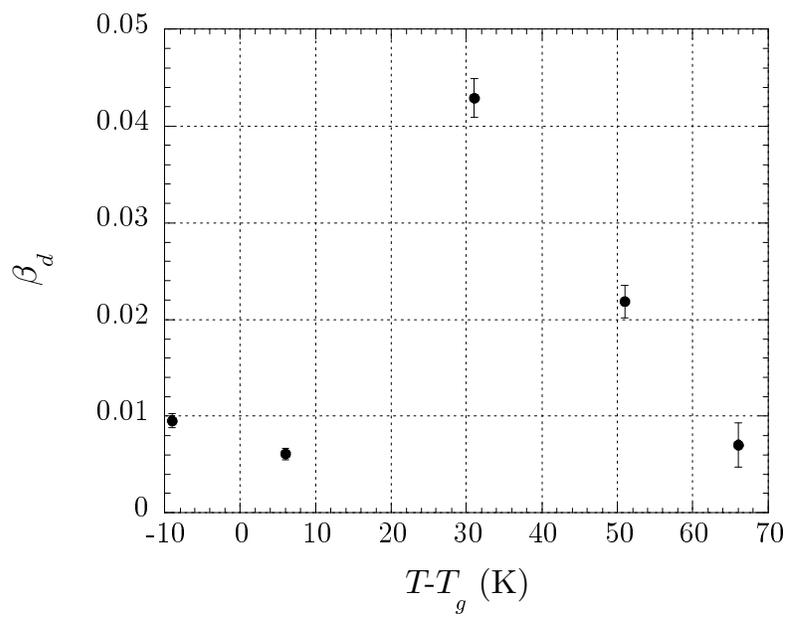
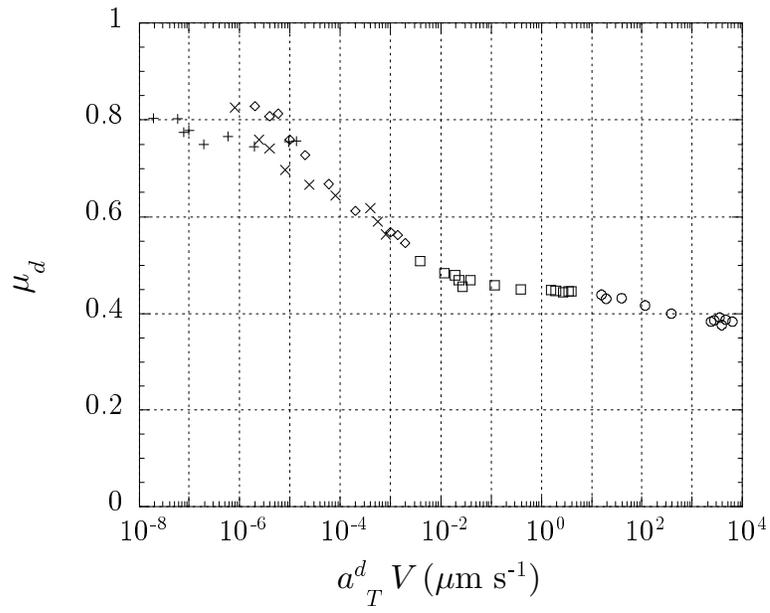
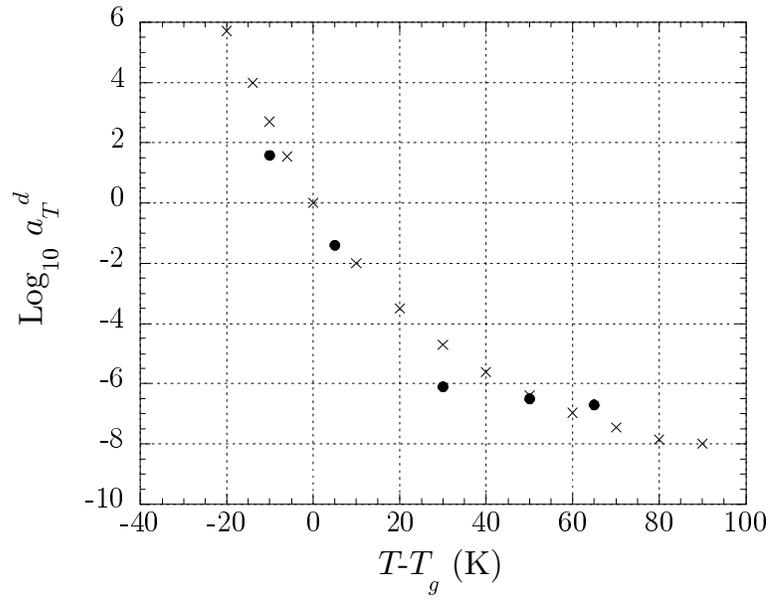
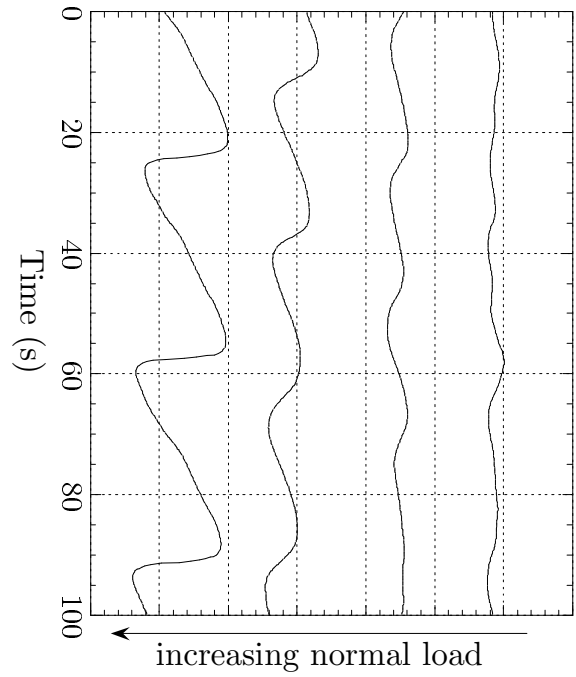
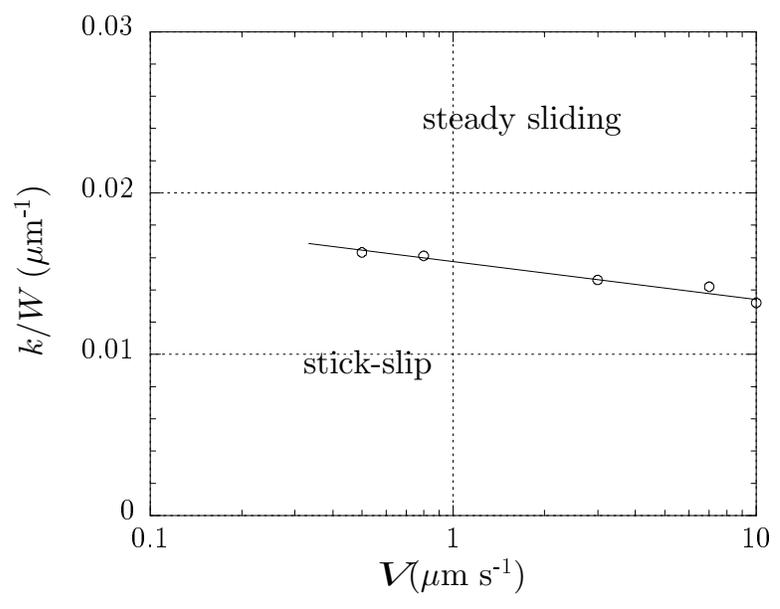
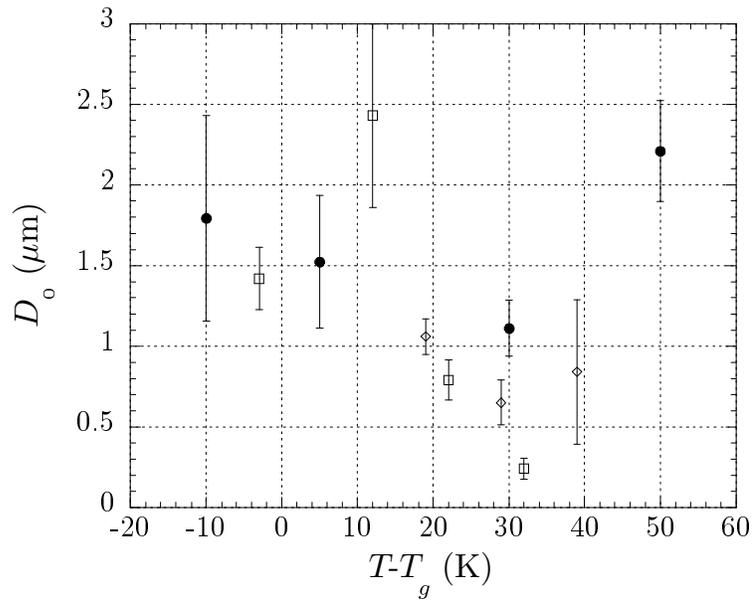
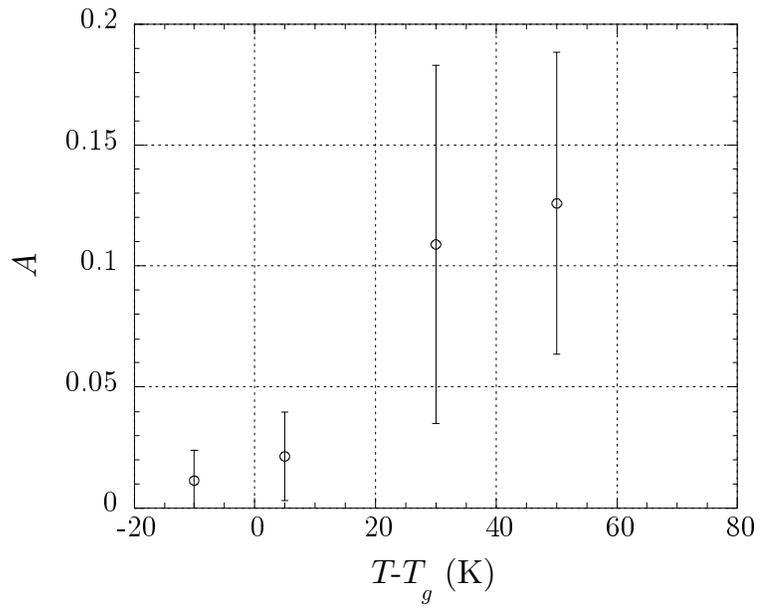
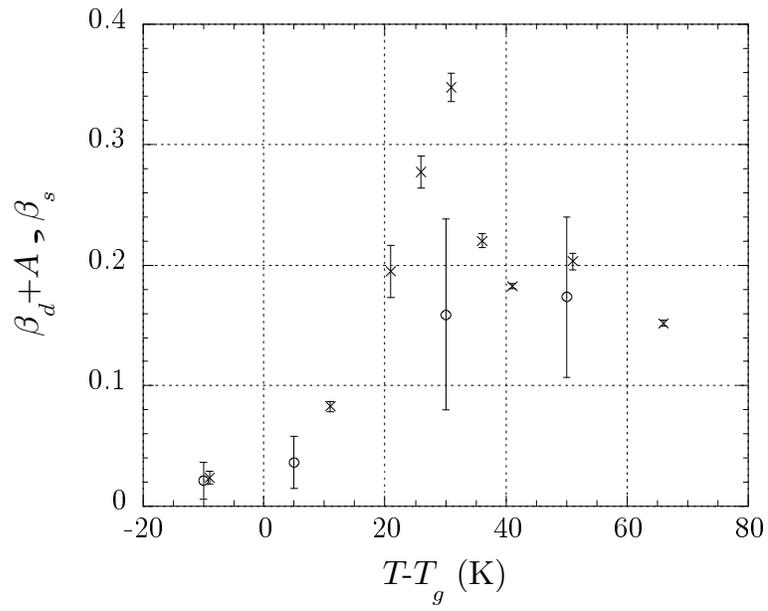
\end{document}